\documentclass{aa}
\usepackage{graphicx,times}

\begin{document}

\title{Cas--A}
\title{X-ray spectral imaging and Doppler mapping of Cassiopeia A}

\author{R.Willingale\inst{1}, J.A.M.Bleeker\inst{2},  
K.J. van der Heyden\inst{2}, J.S.Kaastra\inst{2}, J.Vink\inst{3} }

\offprints{R.Willingale}

\institute{Department of Physics and Astronomy, University of Leicester, 
University Road,
Leicester LE1 7RH
\and
SRON Space Research Institute, Sorbonnelaan 2,
3584 CA Utrecht, The Netherlands
\and
Columbia Astrophysics Laboratory, Columbia University, 550 West 120th Street,
New York, NY 10027, USA
\\email: rw@star.le.ac.uk ; J.A.M.Bleeker@sron.nl ;  K.J.van.der.Heyden@sron.nl
; j.s.kaastra@sron.nl ; jvink@astro.columbia.edu 
\\}

\titlerunning{X-ray Doppler mapping of Cas A}
\authorrunning {R.Willingale et al.}

\date{Received ; accepted }

\abstract{
We present a detailed X-ray spectral analysis of Cas A using
a deep exposure from the EPIC-MOS cameras on-board XMM-Newton. Spectral
fitting was performed on a 15$\times$15 grid of
$20{\arcsec} \times 20{\arcsec}$ pixels using a two component
non-equilibrium ionisation model (NEI)
giving maps of ionisation age, temperature, interstellar column density,
abundances for Ne, Mg, Si, S, Ca, Fe and Ni and Doppler velocities
for the bright Si-K, S-K and Fe-K line complexes.
The abundance maps of Si, S, Ar and Ca are strongly correlated.
The correlation is particularly tight between Si and S. The
measured abundance ratios are consistent with the
nucleosynthesis yield from the collapse of a progenitor star of
12 $M_{\odot}$ at the time of explosion.
The distributions of the abundance ratios Ne/Si, Mg/Si, Fe/Si and Ni/Si
are very variable and distinctly different from
S/Si, Ar/Si and Ca/Si.  This is also expected from the current models of
explosive nucleosynthesis.
The ionisation age and temperature of both the hot and cool NEI
components varies considerably over the remnant. Accurate determination
of these parameters has enabled us to extract reliable Doppler
velocities for the hot and cold components.
The combination of radial positions in the plane of the sky
and velocities along the line of sight have been used to measure
the dynamics of the X-ray emitting plasma. The data are
consistent with a linear radial velocity field for the plasma within the
remnant with $v_{s}=2600$
km s$^{-1}$ at $r_{s}=153$ arc seconds implying a primary shock
velocity of $4000\pm500$ km s$^{-1}$ at this shock radius.
The Si-K and S-K line emission from the cool plasma component
is confined to a relatively narrow shell with
radius 100-150 arc seconds. This component is almost certainly
ejecta material which has been heated by a combination of the
reverse shock and heating of ejecta clumps as they
plough through the medium which has been pre-heated by the primary shock.
The Fe-K line emission is expanding
somewhat faster and spans a radius range 110-170 arc seconds.
The bulk of the Fe emission is confined to two large clumps and it
is likely that these too are the result of ablation
from ejecta bullets rather swept up circumstellar medium.
}

\maketitle

\section{Introduction}

\begin{figure}[!htb]
\centering
\resizebox{\hsize}{!}{\includegraphics{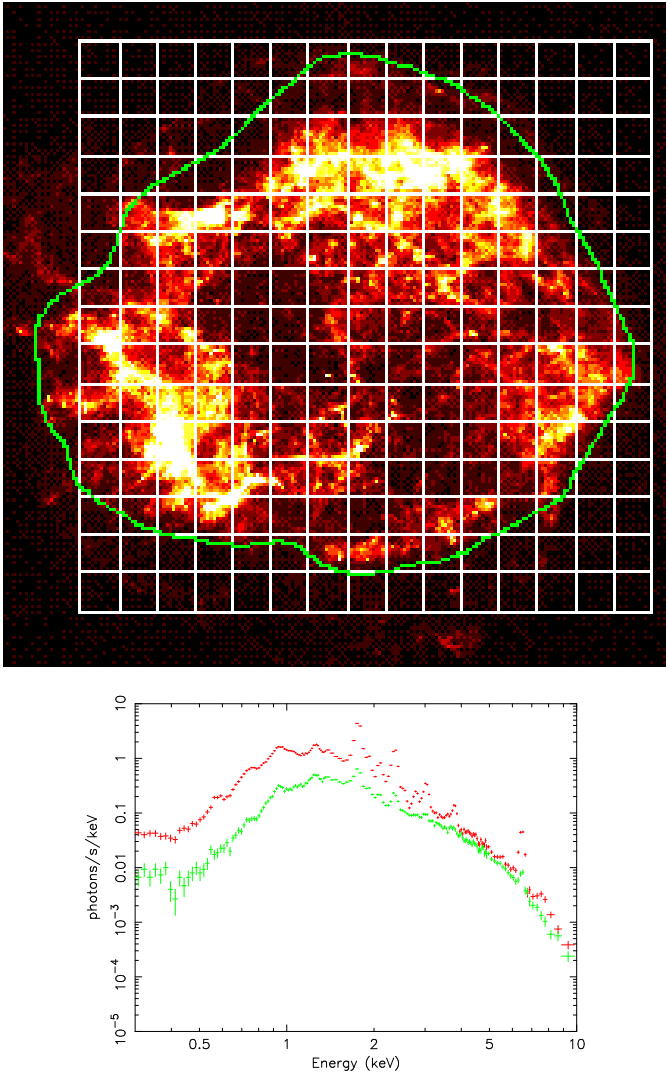}}
\rotatebox{270}{\includegraphics[width=5cm]{fig1b.eps}}
\caption{The pixel grid used in our analysis superimposed on the
high angular resolution Chandra image of Cas A.
The green contour indicates the region
with good statistics and low scattering. Below are typical single pixel
spectra from a high count and a low count region.}
\label{fig1}
\end{figure}

In this paper we present a detailed X-ray spectral analysis of the young 
supernova remnant Cassiopeia A with an angular resolution of the order 20 arc 
seconds over a field of view covering the full remnant.
The data were obtained from an 86 kilosecond 
exposure of the XMM-Newton EPIC-MOS cameras to the source. The outstanding 
spectral grasp of XMM-Newton, i.e. the combination of sensitivity,
X-ray bandwidth 
and spectral resolving power, coupled to this very long exposure time provides 
ample photon statistics for a full spectral modelling of each image pixel 
commensurate with the beam width of the XMM-Newton telescopes ($\sim$15
\arcsec\ Half 
Power Width), even for source regions of low surface brightness. This is 
illustrated in Fig. \ref{fig1} which shows
a broad band high resolution Chandra 
image of Cas A (Hughes et al. 2000)
on which the pixel grid used in this analysis has been 
superimposed. Also drawn on this image is a contour indicating the region with 
good statistics and where the flux is
not dominated by scattering. In addition, two samples of raw 
spectral data are shown, indicating the typical statistical quality in
regions of high and low surface brightness. 

The energy resolution, gain stability and gain uniformity of the MOS-cameras 
allows significant detection of emission line energy shifts of order 1 eV or 
greater for prominent lines like Si-K, S-K and Fe-K.
Proper modelling of these line 
blends with the aid of broad band spectral fitting, taking into account the 
non-equilibrium ionisation balance (NEI), allows an assessment,
with unprecedented accuracy, of Doppler shifts and abundance variations
of the X-ray emitting
material across the face of the remnant with an angular resolution adequate
enough to discriminate the fine knot structure seen by Chandra.
The implications for the dynamical model of the remnant and
for the origin and shock 
heating of the X-ray emitting ejecta will be highlighted as the key
result of this investigation. 

\section{Spectral Fitting Analysis}

We divided a 5\arcmin $\times$ 5 \arcmin field of view of Cas A on
a spatial grid 
containing 15$\times$15 pixels.
This corresponds to a pixel size of $20{\arcsec} 
\times 20{\arcsec}$, slightly larger than the half-power beam width
of XMM-Newton. 
Spectra were extracted using this grid and analysed on a pixel by pixel basis.

The spectral analysis was performed using the SRON SPEX (Kaastra et al. 
\cite{kaastra}) package, which contains the MEKAL code
(Mewe et al. \cite{mewe}) 
for modeling thermal emission. We find that, even at the $20{\arcsec} \times 
20{\arcsec}$ level, one thermal component does not model the data sufficiently 
well, particularly in describing both the Fe-L and Fe-K emission.
We therefore 
choose as a minimum for representative modelling two 
NEI components for the thermal emission. In addition we incorporated the 
absorption measure as a free parameter and also introduced two separate
redshift parameters, one for each plasma component.
 
The basic rationale behind a two component NEI model is that we expect low
and high temperature 
plasma associated with a reverse shock and a blast wave respectively.
While we 
obtain good fits using a two NEI model, we estimate that a contribution from a 
power law hard tail to the 4-6 keV continuum could be as high as 25$\%$. Since 
there is no evidence that the hard X-ray emission is synchrotron and it's 
brightness distribution is very much in line with the thermal component
(see Bleeker et al. \cite{bleeker}),
we feel that our fitting procedure is justified. In other words the combined
high and low 
temperature NEI components will provide a good approximation to the physical 
conditions that give rise to the line emission. 

The low temperature plasma component in our model implicitly assumes that the 
ejecta material, which largely consist  of oxygen and its burning products 
(Chevalier \& Kirshner \cite{chevalier}), has been fully mixed regarding the 
contributing atomic species.
In order to mimic a hydrogen deficient, oxygen rich
medium we adopted a similar approach to that
used by Vink et al. (1996), where
they fixed the oxygen abundance of the cool component to a high value.
We set the 
cool component abundances of O, Ne, Mg, Si, S, Ar and Ca to a factor
10000 higher 
than that of the hot component. It should be noted that 10000 is not
a magic number, 1000 would suffice. The important point is that oxygen
and the heavier elements are all dominant with respect to hydrogen so that
oxygen rather than hydrogen is the prime source of free electrons in the plasma.
The abundances of O, Ne, Mg, Si, S, Ar, Ca, Fe and 
Ni were allowed to vary over the remnant while the rest of the 
elemental abundances (He, C and N)
were fixed at their solar values (Anders \& Grevese \cite{anders}).
 
Our model allows us to estimate the distribution over the remnant of the
emission 
measure $n_{\rm e}n_{\rm H}V$, the electron temperature $T_{\rm e}$ and the 
ionisation age $n_{\rm e}t$ of the two NEI components as well as the
distribution of the abundance of the elements
(O, Ne, Mg, Si, S, Ar, Ca, Fe \& 
Ni), the column density $N_{\rm H}$ of the absorbing foreground material,
Doppler broadening of the lines
and the redshift of the respective plasma components.
Here $n_{\rm e}$ and $n_{\rm H}$ are 
the electron and hydrogen density respectively,
$V$ is the volume occupied by the 
plasma and $t$ is the time since the medium has been shocked.
The best fit model 
parameters were found and recorded for each pixel and it was thus  possible to 
create maps of the various model parameters over the face of the remnant.

\section{Doppler mapping}

It is possible to accurately
determine the Doppler shifts of Si-K, S-K and Fe-K since these lines 
are strong and well resolved.
Doppler shifts of these lines have been calculated in two different ways.

After fits were made to the full spectrum we froze all the fit parameters.
We selected the Si-K (1.72-1.96 keV), S-K (2.29-2.58 keV)
and Fe-K (6.20-6.92 keV) bands for determining their respective
Doppler velocities while ignoring all other line emission.
We then do a fit to each line separately by starting from
the full fit model parameters as a template and subsequently allowing only the
redshift and the abundance of the relevant element to vary.
This method provides a fine tuning of the redshift which in turn
gives the Doppler velocity of the element under scrutiny.

Alternatively, using the same fit parameters we calculated the predicted
continuum flux, $\rm Fp_{cont}$, and energy centroid of the lines,
$\rm Ep_{line}$ and energy centroid of the continuum, $\rm Ep_{cont}$,
for each line energy band in each pixel.
Then using the raw events we calculated the measured total flux,
$\rm F_{band}$ and energy centroid 
$\rm E_{band}$ in each band again for each pixel.
The measured line flux
was then estimated by subtracting the predicted continuum
from the total flux in each band, $\rm F_{line}=F_{band}-Fp_{cont}$.
The continuum centroid in a line energy
band varies as a function of position over the remnant and
the energy centroid for each band is the weighted sum of the
line and continuum components.
The measured line energy centroid was calculated by removing the contribution
from the continuum.
\[E_{line}=(F_{band}\times E_{band}-Fp_{cont}\times Ep_{cont})/F_{line}\]
An estimate of the Doppler shift of the line is then given by
$\Delta E=E_{line}-Ep_{line}$.

The Doppler shifts calculated by these two methods were in reasonable agreement
indicating that the results were not sensitive to line broadening and
lines close to the edges of the chosen energy bands.

\begin{table}
\centering
\begin{tabular}{|l|llll|ll|}
\hline
line & $\rm Ep_{band}$ & $\rm \Delta E_{N}$ & $\rm \Delta E_{C}$ &
$\rm \Delta V$ & $\rm \Delta E_{NEI}$ & $\rm \Delta V_{NEI}$ \\
     &    keV      &    eV          &     eV         &
km s$^{-1}$ & eV & km s$^{-1}$\\
\hline
Si-K    & 1.847 & 0.55  & 0.53 & 94 & 3.89 & 630 \\
S-K   & 2.439 & 0.31  & 0.53 & 99 & 2.52 & 310 \\
Fe-K   & 6.566 & 2.95  & 1.60 & 143 & 24.2 & 1115 \\
\hline
\end{tabular}
\caption[]{Estimates of the mean errors associated with determining the
Doppler velocities.  $\rm \Delta V$ is the mean statistical error in the
Doppler velocity estimated from the two dominant factors
and $\rm \Delta V_{NEI}$ is the rms of the systematic
correction provided by the NEI modelling.}
\label{tab1}
\end{table}
The accuracy of the Doppler shift values depends on the ability
of the NEI spectral model to predict the line blends
combined with the uncertainties in the gain calibration of the detectors.
Maps were constructed using the combined
counts from the two cameras MOS 1 and MOS 2. The results were the same
but with somewhat poorer statistics if only MOS 1 or MOS 2 were used.
The readout
directions of the central CCDs of MOS 1 and MOS 2 are set perpendicular
to one another on the sky so we are sure that the line shifts
are not due to systematic charge transfer losses as the events are
read out. Thus there is convincing evidence
that the observed line shifts are not due to instrumental
gain variations or some other more subtle detector effect.

The statistical errors on the velocity estimates
depend on the intrinsic energy resolution of the detectors, the
number of counts detected in the line energy band and the
error associated with estimating the continuum contribution in the
line energy band. There are two dominant factors. Firstly the statistical
error in determining the energy centroid
$\Delta E_{N}\approx \sigma_{E}/\sqrt(N)$ where $\sigma_{E}$ is the
rms width of the detector energy resolution at the line energy
and $N$ is the total detected count. Secondly the statistical errors
in the continuum bands which translate into errors in the temperature
and flux determination for the continuum flux and hence yield
a centroid error $\Delta E_{C}$.

Systematic errors on the Doppler velocities are introduced by improper
modelling of the unresolved emission line blends within the Si-K, S-K and
Fe-K line energy bands. The centroids of the line blends vary considerably
over the remnant because of large differences in temperature and ionisation
age. We have calculated the rms variation $\rm \Delta E_{NEI}$ of the
band centroid $\rm Ep_{band}$ over the remnant.
If the spectral modelling is correct
then this systematic error should have been eliminated.

Estimates of the factors effecting the accuracy of the Doppler
velocities over the remnant are given in Table \ref{tab1}.
The velocity error $\rm \Delta V$ was calculated from the combination of the two
statistical components.
The velocity errors do vary over the face of the remnant since they
depend on the surface brightness but for all the bright knots the
errors in Table \ref{tab1} are a good estimate.

\section{Key results}

\begin{figure}[!htb]
\centering
\rotatebox{270}{\includegraphics[width=6cm]{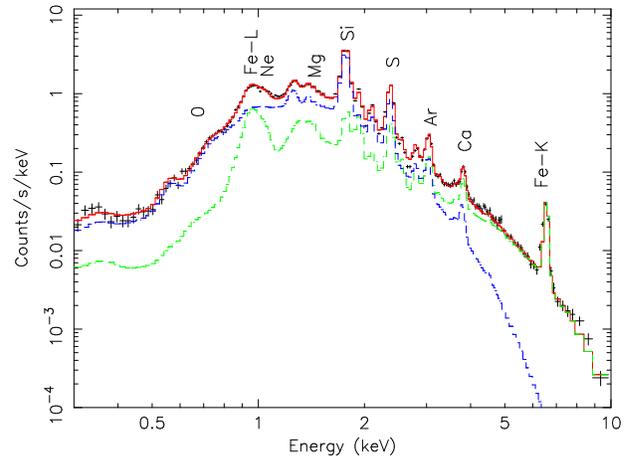}}
\caption{An example of a spectral fit within a single 
$20{\arcsec} \times 20{\arcsec}$ pixel - cool component in blue, hot component
in green and full model in red.}
\label{fig2}
\end{figure}

\begin{figure}[!htb]
\centering
\resizebox{\hsize}{!}{\includegraphics{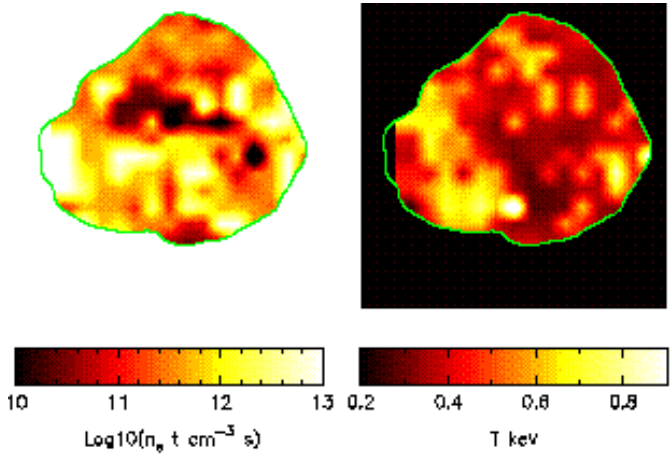}}
\caption{Spectral fit parameters for the cool component, ionisation age
left-hand panel and temperature right-hand panel. The contour indicates
the region with good statistics and low scattering.}
\label{fig3}
\end{figure}

\begin{figure}[!htb]
\centering
\resizebox{\hsize}{!}{\includegraphics{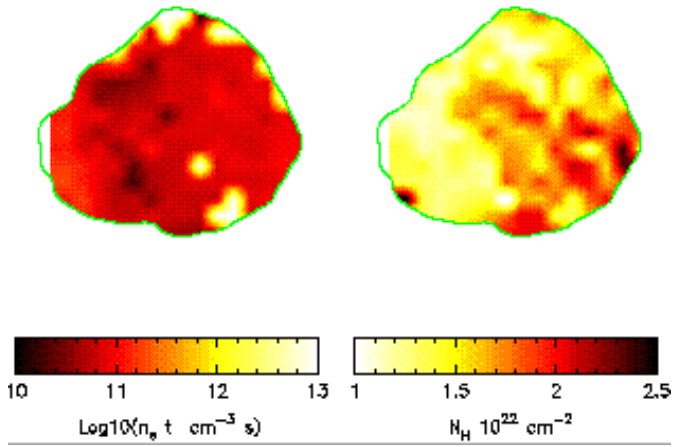}}
\caption{Ionisation age of the hot component and the interstellar
column density. The contour indicates
the region with good statistics and low scattering.}
\label{fig4}
\end{figure}

Fig. \ref{fig2} shows a typical spectral fit. All features of the
measured spectrum are remarkably well represented by the modelling.
Bivariate linear interpolation was used to transfer the
model parameters, predicted fluxes etc. onto the grid of 1 arc second
pixels.
Fig. \ref{fig3} displays maps of the ionisation age and temperature of
the cool NEI component. This component is dominant in the line spectrum
including Fe-L emission. The temperature distribution of the hot
component is similar to (but not the same as) the cool component
but with a temperature
range 2-6 keV. The hot component is responsible for all the
Fe-K emission and also dominates the continuum above 4 keV.
Fig. \ref{fig4} shows the ionisation age of
the hot NEI component and the interstellar column density.
These plots demonstrate the amazing variability in the spectrum over
the face of the remnant.
The column density does exhibit some correlation with the surface
brightness of the bright knots of the remnant presumably because of
parameter coupling in the spectral fitting process. The $\rm N_{H}$
fitting is
particularly sensitive to modelling of the
O VIII emission 0.6-0.8 keV and Fe-L lines 1.0-1.5 keV.
The mean column
density is $1.5\times10^{22}$ cm$^{-2}$ while the range is
$1.0-2.5\times10^{22}$ cm$^{-2}$. The variation of interstellar
column density over the face of the remnant has previously been mapped by
Keohane et al. (1996) using radio data.
The distribution in their map is similar to Fig. \ref{fig2} with
high column density in the West due to a molecular cloud
but the overall range of their column density derived from the equivalent
widths of HI and OH is smaller, $1.05-1.26\times10^{22}$ cm$^{-2}$.

\begin{figure}[!htb]
\centering
\resizebox{\hsize}{!}{\includegraphics{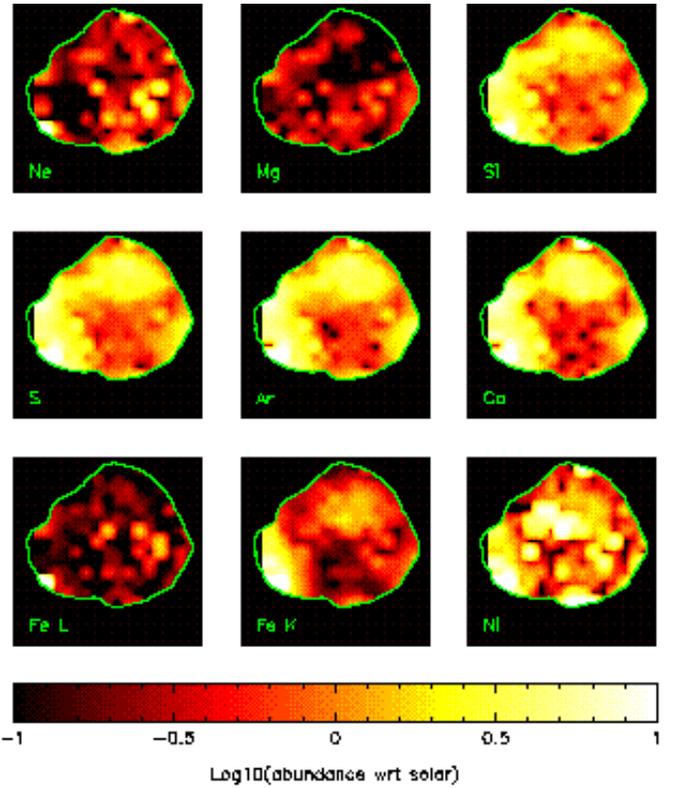}}
\caption{Abundance maps for the elements included in the spectral fitting.
All are plotted on the logarithmic scale indicated by the bar at the 
bottom.}
\label{fig5}
\end{figure}

\begin{figure}[!htb]
\centering
\includegraphics{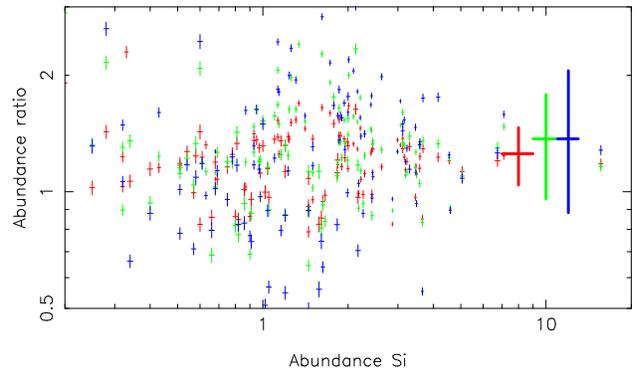}
\caption{The variation in the abundance ratios of S/Si red, Ar/Si green 
and Ca/Si blue as a function of the Si abundance. The large error bars
to the right indicate the mean and rms scatter for the three elements.}
\label{fig6}
\end{figure}

Fig. \ref{fig5} is a montage of abundance maps. Again we see
considerable variations over the remnant. The Fe-L distribution comes
from the cool component while the Fe-K and Ni are derived
exclusively from the hot component.
The distributions of Si, S, Ar
and Ca, which are all oxygen burning products, are similar and
distinct from carbon burning products, Ne and Mg, and Fe-L.
Fig. \ref{fig6} shows the variation in the ratios S/Si, Ar/Si and S/Si
with respect to the abundance of Si.
On the one hand these ratios clearly
vary over the remnant but on the other hand, for a Si abundance range spanning
more than two orders of magnitude, these ratios remain remarkably constant.
The thick vertical bars indicate the mean and rms scatter of the
ratio values.

\begin{figure}[!htb]
\centering
\resizebox{\hsize}{!}{\includegraphics{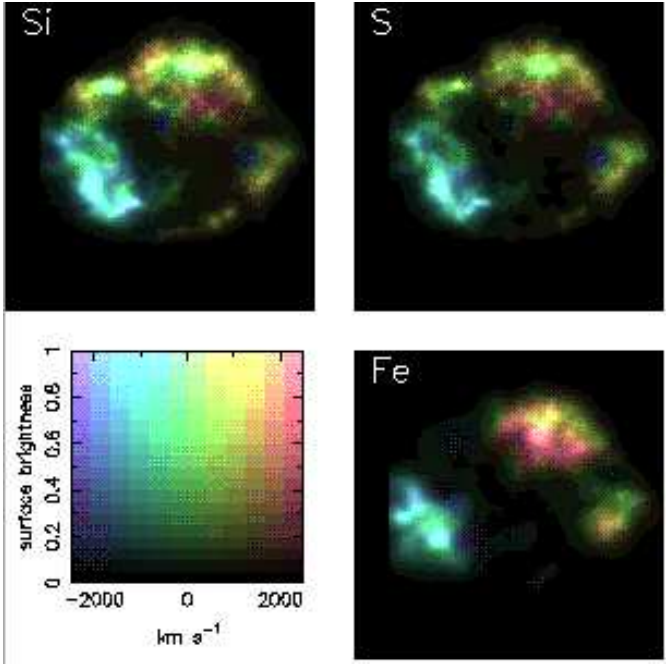}}
\caption{Doppler maps derived from Si-K, S-K and Fe-K emission lines.
For each case the surface brightness of the line emission (after
subtraction of the continuum) is shown
colour coded with the Doppler velocity. The coding used
is shown in the bottom left image.}
\label{fig7}
\end{figure}

\begin{figure}[!htb]
\centering
\includegraphics{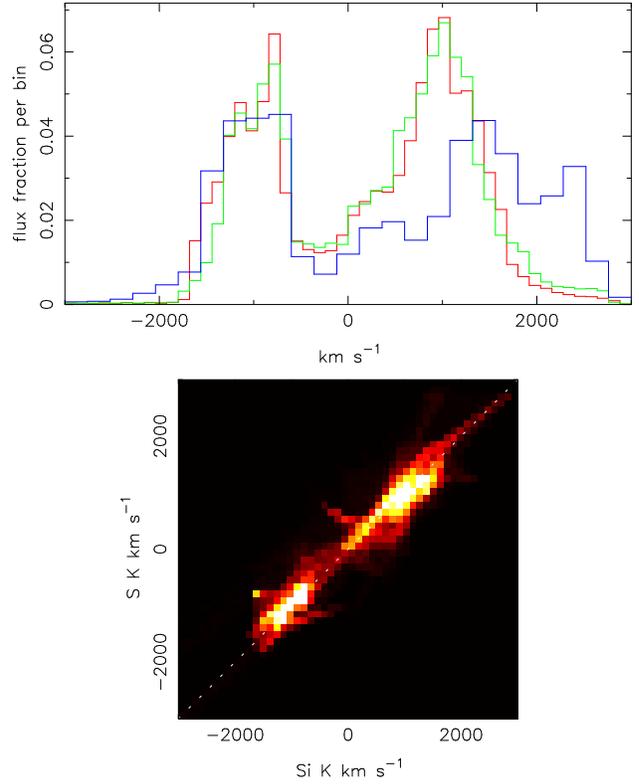}
\caption{Flux distributions of Si-K (red), S-K (green)
and Fe-K (blue) as a function of measured Doppler velocity.
The lower panel shows the flux distribution in the Si velocity-S velocity
plane.}
\label{fig8}
\end{figure}

Line flux images were produced using an adaptive filter with a minimum
beam count of 400 and maximum beam radius of 15 arc seconds. The
raw event images from each line energy band were smoothed and then
multiplied by the ratio of the predicted line flux to line plus continuum flux
ratio in order to estimate the line flux.
Fig. \ref{fig7} shows the resulting line flux images colour
coded with the Doppler velocity.
The bottom left image is the colour coding used.
The Doppler shifts seen in different areas of the remnant are very
similar in the three lines. The knots in the South East are blue
shifted and the knots in the North are red shifted.
This is consistent with previous measurements, Markert et al. (1983),
Holt et al. (1994), Vink et al. (1996).
Moving from large radii towards the centre the shift generally gets larger
as expected in projection. This is particularly pronounced in the North.
At the outer edges the knots are stationary or slightly blue shifted.
Moving South a region of red shift is reached indicating these inner
knots are on the far side of the remnant moving away from us.
The distributions of flux as a function of Doppler velocity are
shown in Fig. \ref{fig8}. The distibutions for Si-K and S-K are very
similar. The Fe-K clearly has a slightly broader distribution for the
red shifted (+ve) velocities. The lower panel is the flux plotted
in the Si velocity-S velocity plane showing the tight correlation between
the Doppler shift measured for these lines. This plot was quite sensitive to
small systematic changes in temperature, ionisation age or abundances
in the spectral fitting since these can potentially have a
profound effect on the derived Doppler velocities as indicated by
the large values of $\rm \Delta V_{NEI}$ in Table \ref{tab1}.

The X-ray knots of Cas A form a ring because the emitting
plasma is confined to an irregular shell. We searched for
a best fit centre to this ring looking for the position that gave the
most strongly peaked radial brightness distribution (minimum
rms scatter of flux about the mean radius). The best centre
for the combined Si-K, S-K and Fe-K line image was 13 arc second West
and 11 arc seconds North of the image centre (the central Chandra point source).
Using this centre the peak flux occured at a radius of 102 arc secs,
the mean radius was 97 arc secs and the rms scatter about the
mean radius was 24 arc seconds.

\begin{figure}[!htb]
\centering
\includegraphics{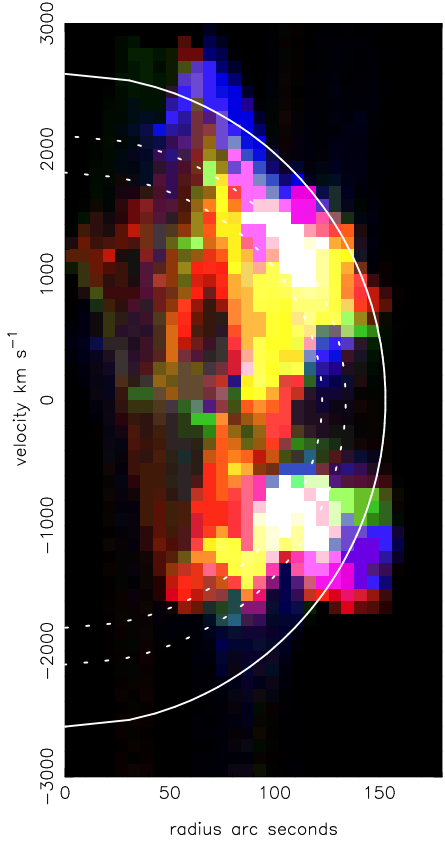}
\caption{Flux distribution of Si-K (red), S-K (green) and Fe-K (blue) in the
radius-velocity plane. The solid line is the best fit shock radius (see
text). The outer
dotted line indicates the peak of the Fe-K flux distribution and
the inner dotted line indicates the mean radius of the Si-K and S-K
emission.}
\label{fig9}
\end{figure}

Given such a centre we can assign a radius to each pixel and using
the Doppler velocity measured for each pixel we can map the flux
into the radius-velocity plane. The result is shown in Fig. \ref{fig9}.

\section{Ionisation structure and abundance ratios}

The models for nucleosynthesis yield from massive stars predict that the
mass or abundance ratio $R_{\rm X/Si}$ of ejected mass of any element X
with respect to silicon varies significantly as a function of the
progenitor mass $M$. We show the observed mean values of $R_{\rm X/Si}$
as well as its rms variation in Table \ref{tab2}, together with the predictions
for models with a progenitor mass of 11, 12 and 13 $\rm M_{\odot}$.
\begin{table}
\centering
\begin{tabular}{|l|ll|lll|}
\hline
ratio & mean & rms & 11 $\rm M_{\odot}$ & 12 $\rm M_{\odot}$ & 13 $\rm 
M_{\odot}$ \\
     \hline
O/Si    & 1.69 & 1.37  & 0.44 & 0.16 & 0.33 \\
Ne/Si    & 0.24 & 0.37 & 0.59 & 0.12 & 0.33 \\
Mg/Si    & 0.16 & 0.15 & 0.57 & 0.12 & 0.41 \\
S/Si    & 1.25 & 0.24 & 0.87 & 1.53 & 0.88  \\
Ar/Si    & 1.38 & 0.48 & 0.65 & 2.04 & 0.64 \\
Ca/Si    & 1.46 & 0.68 & 0.63 & 1.62 & 6.56 \\
FeL/Si    & 0.19 & 0.65 & 1.37 & 0.23 & 0.96 \\
FeK/Si    & 0.60 & 0.51 & 1.37 & 0.23 & 0.96 \\
Ni/Si    & 1.67 & 5.52 &  6.89 & 0.68 & 1.80 \\
\hline
\end{tabular}
\caption[]{Mean measured abundance ratios and rms scatter compared with
theoretical predictions for progenitor masses of 11, 12 and 13 $\rm
M_{\odot}$.}
\label{tab2}
\end{table}
The observed abundance ratios for X equal to Ne,
Mg, S, Ar, Ca and Fe-L (the iron of the cool component) are all
consistent with a progenitor mass of 12.0$\pm$ 0.6 $\rm M_{\odot}$,
where we used
the grid of models by Woosley and Weaver (1995).  In these models, most
of the Si, S, Ar and Ca comes from the zones where explosive O-burning
and incomplete explosive
Si-burning occurs, and indeed as Fig. \ref{fig6}
shows these elements track each
other remarkably well. Furthermore the average abundance ratio fits the
expectation for a 12 $\rm M_{\odot}$ progenitor.
The correlation between Si and
S is remarkably sharp but not perfect, the scatter in Fig. \ref{fig6}
is real. These remaining residuals can be attributed to small
inhomogeneities. Table \ref{tab2} also indicates that the rms
scatter of the abundances with respect to Si get larger as Z increases,
S to Ar to Ca and indeed through to Fe and Ni. This is to be expected since
elements close together in Z are produced in the same layers within
the shock collapse structure while those of very different Z are
produced in different layers and at different temperatures.

The Fe which arises from complete and incomplete Si burning should give
rise to iron line emission. For both the Fe-L and Fe-K lines we see that
iron abundance varies over the remnant but does not show any
straightforward correlation with the other elements (there is a very
large scatter in $R_{\rm Fe/Si}$).
This is to be expected if most of the iron arises
from complete Si burning. We return to the different morphologies of Si and Fe
later in the discussion.

Ne and Mg are mostly produced in shells where Ne/C burning
occurs, and the relative scatter in terms of $R_{\rm X/Si}$ is indeed
much larger than for S, Ar and Ca (Table \ref{tab2}).
Furthermore the abundance maps of Ne and Mg in Fig. \ref{fig5} are
similar and very different from the Si, S, Ar and Ca group.

The oxygen abundance is much higher than predicted by theory, contrary to
all other elements. We cannot readily offer an explanation for this, but
there are at least two complicating factors.
As the XMM RGS maps show (Bleeker et al.
2001), oxygen has a completely different spatial distribution to the
other elements (it is more concentrated to the North), and it is also much
harder to measure due to the strong galactic absorption and relatively
poor spectral resolution of the EPIC cameras at low energies.

The map of the ionisation age of the cool component shows a large
spread.  The average value at the Northern rim (few times
$10^{11}$~cm$^{-3}$s) matches nicely the value derived from ASCA data
(Vink et al.  1996).  At the SE rim the ionisation age is much larger
(cf.  Vink et al.  $4\times 10^{11}$~cm$^{-3}$s).  We confirm this higher
value, but also see that there is a large spread in ionisation age.  It
should also be noted that for ionisation ages larger than about
$10^{12}$~cm$^{-3}$s the plasma is almost in ionisation equilibrium and
therefore the spectra cannot be distinguished from equilibrium spectra;
the extremely high values of $10^{13}$~cm$^{-3}$s in the easternmost part
of the remnant (Fig. \ref{fig3}) are therefore better interpreted as being just
larger than $10^{12}$~cm$^{-3}$s.  There is also a region of very low
ionisation age (less than $3\times 10^{10}$~cm$^{-3}$s) stretching from East
to West just above the centre of the remant.  This region also has a very
low emissivity (i.e. low electron density)
and can be understood as a low density wake just behind
and inside of the shocked ejecta.

The hot component has a more homogeneous distribution of ionisation age,
centered around $10^{11}$~cm$^{-3}$s, again consistent with the typical
value found by Vink et al. (1996) but in that case integrated over much
larger areas. We have now clearly resolved this component spatially.

\section{Dynamics}

\begin{figure}[!htb]
\centering
\includegraphics{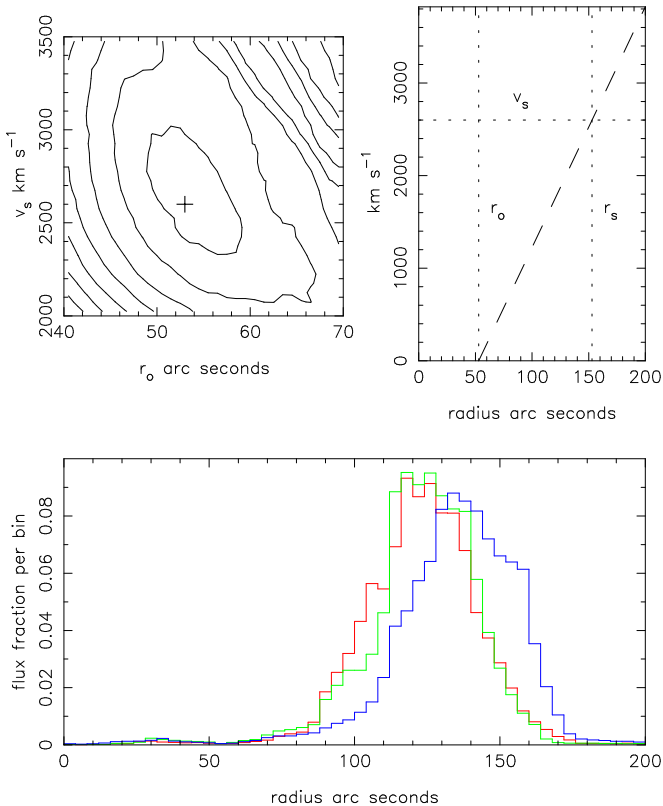}
\caption{The velocity field which gives the minimum normalised shell thickness
$\rm \Delta r/r$. Each contour interval in the top left plot
corresponds to a $\sim2$\% increase in shell thickness.
The lower panel shows the deprojected
flux distribution of Si-K (red), S-K (green) and Fe-K (blue)
as a function
of radius from the centre of the remnant.}
\label{fig10}
\end{figure}

In the radius-velocity plane the flux from a thin shell of radius
$\rm r_{s}$ expanding at velocity $\rm v_{s}$ is
expected to form an ellipse which intersects the radius axis at $\rm r_{s}$
and the velocity axis at $\rm \pm v_{s}$. We expect the velocity
to increase with radius and for simplicity we can assume that the
velocity field within the spherical volume is given by a linear form

\[\rm v(r)=\frac{v_{s}}{(r_{s}-r_{o})}(r-r_{o})\]

where $\rm v_{s}$ is the postshock velocity of material
at the shock radius $\rm r_{s}$
and $r_{o}$ is the radius within the remnant
at which the velocity falls to zero.
From the analysis of the
Chandra data, (Gotthelf et al. 2001), we have an estimate of $\rm r_{s}=153$
arc seconds. Adopting such a spherically symmetric velocity field 
precludes the possibility that the expansion is wildly asymmetric.
In the plane of the sky the outer shock seen both in radio and X-ray images
is remarkably circular (see for example the Chandra X-ray image
scaled to highlight the faint outer filaments, Gotthelf et al. 2001).
We therefore think it is unlikely that the velocity field or
shock radius is very different along the line of sight.
Assuming values for $\rm v_{s}$ and $\rm r_{o}$ we can
calculate the radius of each pixel within the spherical volume
from the observed radius on the plane of the sky and the Doppler
velocity (along the line of sight). As the parameters are changed
so the fractional rms scatter, $\rm \Delta r/r$, of the flux about
the mean radius varies. The top left panel of
Fig. \ref{fig10} shows the variation
in $\rm \Delta r/r$ as a function of $\rm r_{o}$ and $\rm v_{s}$.
The minimum scatter of 0.16 is found at $\rm r_{o}=53$ arc seconds
and $\rm v_{s}=2600$ km s$^{-1}$. The velocity field which gives the
minimum shell thickness is shown in the top right-hand panel.
This result is very similar to the velocity field
of an isothermal blast wave, Solinger et al. (1975), but we would like to
stress this does not imply that Cas A has entered this phase of it's
evolution. The remnant is almost certainly in between the ejecta-dominated
and the Sedov-Taylor stages as modelled in detail by Truelove and
McKee (1999).
The lower panel of Fig. \ref{fig10} shows the deprojected
flux distributions as a function of radius in
the spherical volume. The Si-K and S-K distribution match quite closely.
The Fe-K distribution is somewhat broader and peaks at a larger radius.
The deprojected Si-K and S-K flux distributions
are very similar to the emissivity profiles derived from the Chandra data
(Gotthelf et al. 2001) although their analysis was for a section
of the remnant while the profiles in Fig. \ref{fig10} are
the average over the full azimuthal range.
The solid semi-circular
line in Fig. \ref{fig9} is the best fit with $\rm r_{s}=153$
arc seconds and $\rm v_{s}=2600$ km s$^{-1}$. The inner dotted
line shows the mean radius of the combined Si-K and S-K flux and
the outer dotted line shows the peak of the Fe-K distribution. The outer
reaches of the Fe-K distribution straddle the shock radius derived from
the Chandra image.

The MOS energy resolution cannot separate the red and
blue components when they overlap. If we see both the distant red
shifted shell and nearer blue shifted shell in the same beam the
line profile is slightly broadened but the centroid shift is
diminished. The observed Doppler velocities and the best
fit value for $\rm v_{s}$ may be slightly biased by this ambiguity, however
most beams appear to be dominated
by either red or blue shifted knots and therefore this bias is
expected to be small. It is fortuitous that the X-ray emission is
distributed in clumps rather than a thin uniform shell since this
enables us to measure the Doppler shift with a modest angular
resolution without red and blue components in the same beam 
cancelling each other out.

\begin{figure}[!htb]
\centering
\resizebox{\hsize}{!}{\includegraphics{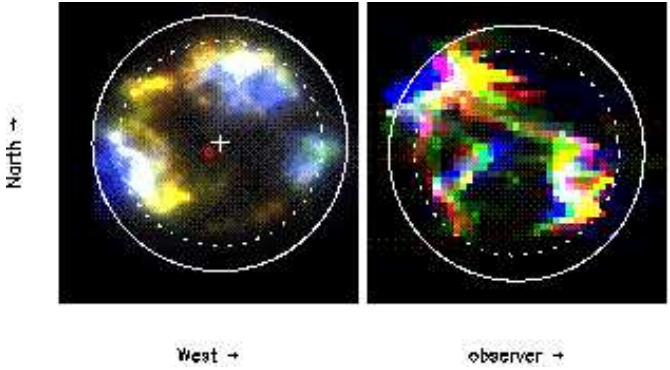}}
\caption{The left-hand panel is an image of Si-K (red), S-K (green)
and Fe-K (blue).
The small red circle indicates the position of the
Chandra point source. The white cross
is the best fit centre from the fitting of the radial distribution.
The right-hand panel is a reprojection of the same line fluxes
onto a plane containing the line of sight, North up, observer to right.
In both panels the outer solid circle is the shock radius
$\rm r_{s}=153$ arc seconds and
the inner circle is the mean radius of the Si-K and S-K flux.}
\label{fig11}
\end{figure}

The left-hand panel of Fig. \ref{fig11} is a composite image of the
remnant seen in the Si-K, S-K and Fe-K emission lines.
The solid circle indicates
the $\rm r_{s}=153$ arc seconds and the dashed circle is
the mean radius of the Si-K and S-K flux $\rm r_{m}=121$ arc seconds.
The X-ray image of the remnant provides coordinates x-y
in the plane of the sky.
Using the derived radial velocity field within the remnant
we can use the measured Doppler velocities $\rm v_{z}$ to give us
an estimate of
the z coordinate position of the emitting material along the line of
sight thus giving us an x-y-z coordinate
for the emission line flux in each pixel. Using these coordinates we
can reproject the flux into any plane we choose. The right-hand panel
of Fig. 11 shows such
a projection in a plane containing the line-of-sight, North upwards,
observer to the right.
In this reprojection the line emission from Si-K, S-K and Fe-K are
reasonably well aligned for the main ring of knots. The reprojection
is not perfect because the MOS cameras are unable to resolve
components which overlap along the line-of-sight and this
produces some ghosting just North of the centre of the remnant.
In the plane of the sky Fe-K emission (blue) is clearly
visible to the East
between the mean radius of the Si+S flux and the shock radius.
Similarly in the reprojection Fe-K emission is seen outside the main
ring in the North away from the observer. The Si+S knot in the South away from
the observer
in the reprojection is formed from low surface brighness emission
in the South West quadrant of the sky image.
The X-ray emitting material is very clumpy within the spherical volume
and is indeed surprisingly well characterised by the doughnut shape
suggested by Markert et al. (1983). However the
distribution is distinctly different to that obtained in similar
3-D studies of the optical knots, Lawrence et al. (1995).

The expansion of Cas A has been measured in various ways;
using the proper motion of optical knots
(van den Bergh and Kamper 1983, Fesen et al. 1987, Fesen et al. 1988),
from the proper motion of radio knots (Anderson and Rudnick 1995),
using Doppler shifts of spectral lines from optical knots (Reed et al. 1995,
Lawrence et al. 1995),
Doppler shift of X-ray line complexes (Markert et al. 1983,
Holt et al. 1994 and Vink et al. 1996) and the proper motion
of X-ray knots (Vink et al. 1999). These methods identify a number
of distinct features with different dynamics; Quasi Stationary Flocculi
(optical QSF), Slow Radio Knots in the South West (SRK), the main
ring of radio knots, the main ring of X-ray knots (continuum + lines
1-2 keV), Fast Moving Knots (optical FMK) and Fast Moving Flocculi
(optical FMF).

In proper motion studies it is conventional to express the motion
as an effective expansion time $\rm t_{ex}=R/V$ (years) where $R$ is the
radius of the feature/knot from some chosen centre (arc seconds) and
$V$ is the proper motion (arc secs/year).
The deceleration parameter, the ratio of the true age over the expansion
age, can be estimated as $\rm m=t_{age}/t_{ex}$. There is no need to
deproject the radius or velocity to estimate $m$. However if
we then wish to estimate a true expansion velocity the R must be
deprojected but still the ratio $R/V$ will remain constant.
Doppler measurements allow some form of deprojection and measured
radii on the sky can be converted to actual radii within the volume
of the remnant as described in the previous section.
Given a radius in arc seconds $r_{as}$ and velocity in km s$^{-1}$
$\rm v_{kms}$ we can calculate an expansion time in years $\rm t_{ex}$ assuming
a distance in kpc $\rm d_{kpc}$,
$\rm t_{ex}=4.63\times10^{3} r_{as}d_{kpc}/v_{kms}$.
Previous authors have used combinations of these measurements
to refine estimates of the age and/or distance. Alternatively
we can adopt some age and distance and compare the radii and expansion
velocities of the various components. The original explosion probably
occured in 1680 (Ashworth 1980) so the age in 2000 is $\rm t_{age}=320$ years.
Distance estimates have varied over the years but recent studies (Reed et
al. 1995) have settled on $3.4^{+0.3}_{-0.1}$ kpc.

Table \ref{tab3} gives estimates of the expansion parameters for the
different components. Those marked with an asterisk are from
proper motion studies which estimate the expansion time or
the deceleration parameter directly. For these
the $\rm r_{as}$ value has been estimated and the $\rm v_{s}$ calculated using
the measured expansion time. From the Doppler measurements we get
a measurement of $\rm r_{s}$ and $\rm v_{s}$ which are then used to estimate
the expansion time or the deceleration parameter.
Proper motion studies of X-ray emission track the movement of
shock features in the plane of the sky while
X-ray emission line Doppler measurements estimate the velocity of the postshock
plasma $\rm v_{s}$ along the line of sight.
The shock velocity $\rm u_{s}$ is related to the postshock plasma velocity,
$\rm v_{s}=\alpha u_{s}$.
The factor $\rm \alpha$ depends on the thermodynamics of the shocked gas
but ranges
between 0.58 for isothermal to 0.75 for
adiabatic conditions, see for example Solinger et al. (1975).
The present X-ray emission line (Xline) results in Table \ref{tab3}
have been calculated from the derived velocity field parameters,
$\rm r_{s}=153$ and $\rm v_{s}=2600$
using a mean value of $\rm \alpha=0.65$. The
error quoted reflects the uncertainty in this factor.
The FMF
are at large radii so it is likely that the deprojection correction
is small and the value of 168 arc secs quoted is in fact
the mean radius in the plane of the sky. For the SRK in
the South West sector
and the QSF the values quoted for $\rm r_{s}$ are just reasonable
guesses.
\begin{table}
\centering
\begin{tabular}{|l|llll|}
\hline
& $\rm v_{kms}$ & $\rm r_{as}$ & $\rm t_{ex}$ & m \\
\hline
QSF* & $370\pm300$ & $\sim105$ & $4470\pm2300$ & $0.07\pm0.06$ \\
SRK* & $656\pm16$ & $\sim100$ &  $2100\pm360$ & $0.15\pm0.02$ \\
Radio*&$2848\pm100$ & $\sim110$ & $604\pm12$  & $0.51\pm0.02$ \\
Xline& $4000\pm500$ & $153\pm5$  & $537\pm70$ & $0.60\pm0.07$ \\
1 keV*&$3456\pm105$ & $\sim110$  & $501\pm15$  & $0.63\pm0.02$ \\
FMK  & $5290\pm90$ & $105\pm1$  &  $312\pm9$  & $0.98\pm0.03$ \\
FMF* & $8816\pm28$ & $168\pm6$  &  $300\pm9$  & $0.99\pm0.03$ \\
\hline
\end{tabular}
\caption[]{The expansion parameters for radio, optical and
X-ray emissions. * indicates proper motion studies. The results
from this paper are given in the line labelled Xline (see text for
details).}
\label{tab3}
\end{table}
The expansion time and deceleration parameters
derived from the observed Doppler shifts of the Si-K, S-K and Fe-K lines
(Xline in Table \ref{tab3})
are in reasonable agreement with the radio proper motion observations
(Anderson and Rudnick, 1995) and in good agreement with the
soft X-ray proper motion measurements of Vink et al. (1999) (1 keV in
Table \ref{tab3}).

\section{Discussion}

The tight correlation between the variation in abundance of Si, S, Ar, Ca
over an absolute abundance range of two orders of magnitude is
strong evidence for the nucleosynthesis of these ejecta elements by
explosive O-burning and incomplete explosive Si-burning due to the shock
heating of these layers in the core collapse supernova.
Full mixing of the burning products is implied by the excellent fit
to the plasma model. However the Fe emission, both in the Fe-K and the Fe-L
lines, does not show this correlation in any sense. A significant fraction
of the Fe-K emission is seen at larger radii than Si-K and S-K as
convincingly demonstrated in our Doppler derived 3-D reprojection,
Fig. \ref{fig11}.
Moreover the Fe-K emission is patchy, reminiscent of large clumps
of ejecta material, rather than shock heated swept up circumstellar material.
In fact the bulk of the Fe-K emission arises in two limited regions
possibly indicating that the core collapse threw off material in two
opposing clumps which we clearly see in Fig. \ref{fig11}.
If we interprete these Fe-rich ejecta as the nucleosynthesis product
of complete explosive burning of the Si-layer, spatial inversion of
the O- and Si-burning products has
occurred
and large scale bulk mixing of the explosion products is an inevitable
consequence.
A similar conclusion was obtained by Hughes et al. (2000) for ejecta
material at the  east side of the remnant based on the morphological
features of the high spatial resolution Chandra data.

The Fe-K emission requires a relatively high temperature in the range 2-6 keV.
This temperature cannot be generated by the reverse shock wave,
but only by the primary blast wave.
Heating is certainly provided by the primary
shock but preheating of the ambient medium by clumps that move ahead
of the primary shock could contribute, see Hamilton (1985).

In the plane of the sky image Fig. \ref{fig11}
the Fe-K emission to the East is at a radius of
140 arc seconds, near the primary shock, with an implied shock velocity of
$\rm u_{s}$ in the range 3500-4500 km s$^{-1}$ (see previous section).
It is coincident with a cluster of three FMFs,
4,5,6 listed by Fesen et al. (1988).  They all have a proper
motion of $0.52\pm0.05$ arc secs yr$^{-1}$ and a mean radius from the
expansion centre in 1976 of $149\pm3$ arc seconds. Assuming
a distance of 3.4 kpc and age in 1976 of 296 years
this corresponds to a transverse velocity of $8170\pm790$ km s$^{-1}$
and a deceleration parameter of $m=1.0\pm0.1$.
The same Fe-K emission is also coincident with the radio knots
89,90,92 and 93 listed by Anderson and Rudnick (1995).
These have a mean proper motion
of $0.26\pm0.01$ arc secs yr$^{-1}$ and a mean radius from the
expansion centre in 1987 of $136\pm3$ arc seconds.
This corresponds to a transverse velocity of $4117\pm160$ km s$^{-1}$
and a deceleration parameter of $m=0.59\pm0.02$.
These radio knots also correspond to the bow shock feature D identified
using morphology and polarimetry by Braun, Gull \& Perley (1987). They
estimate the Mach number of this feature as 5.5, the highest in their
list of 11 such features.

The Fe-K emission at large radii is
highly reminiscent of SNR shrapnel discovered by Aschenbach et al.
(1995) around the Vela SNR. These are almost certainly bullets of
material which were ejected from the progenitor during the
collapse and subsequent explosion. They would initially be expected to
have a radial velocity less than the blast wave but as the remnant
develops, and the shock wave is slowed by interaction with the
surrounding medium, the bullets would overtake the blast wave
and appear outside the visible shock front as is the case
in Vela. It was suggested by Aschenbach et al. (1995) that the X-ray
emission from the Vela bullets arises from shock-heating of
the ambient medium by supersonic motion. If this is the case
the X-rays will be seen from Mach cones which trail the bullets
extending back towards the centre of the remnant.

The generation of radio emission associated with the deceleration of
ejecta bullets has been discussed at length by several authors,
Bell (1977), Braun, Gull \& Perley (1987), Anderson and Rudnick (1995).
The optical emission arises from shocks penetrating dense
ejecta clumps. When these internal shocks have crossed the clump
deceleration sets in accompanied by a strong turn-on of radio
synchrotron emission. Electrons are accelerated in the bow-shock and
the magnetic field is amplified in shearing layers between the
dense ejecta and the external medium. The amplified magnetic
field in the wake of ejecta bullets
is predominately radial in agreement with radio polarization
measurements, Anderson, Keohane and Rundick (1995).
The supersonic flow associated with this scenario has been
simulated by Coleman \& Bicknell (1985).
The same situation could also give rise to X-ray emission. The bulk of
the electrons are heated to $\sim3$ keV by the bow shock. As the
shocked material drifts back into the wake the plasma slowly comes
into ionization equilibrium and X-ray line emission is produced.
Our analysis of the abundances clearly indicates that
the matter responsible for the line emission is ejecta and this must
have been ablated from the bullets rather than swept up by the shock.
The velocities of both the radio and
X-ray emission in the East are about half that of the optical.
This is consistent with the peak of the radio and X-ray emission
falling in the wake of the bullet trailing behind the peak
of the optical emission.

What is the heating mechanism responsible for the cool component? Our
present analysis clearly indicates this component is dominated by ejecta
material. It is conventional to assume that
the primary source of ejecta heating which produces the bright
ring of X-ray emission in Cas A is the reverse shock (McKee 1974,
Gull 1975). However the primary shock seen in X-rays and radio at a radius
of 150 arc seconds is not very bright and it is not clear that the
reverse shock has been or is presently very strong.
The Chandra image shows much fragmentation
consistent with dense bullets and it is likely that significant
heating arises, again, from the interaction of these bullets with
the material pre-heated by the primary shock.

\begin{acknowledgements}
The results presented are based on observations obtained with XMM-Newton,
an ESA science
mission with instruments and contributions directly funded by
ESA Member States and the USA.
JV acknowledges support in the form of the NASA Chandra Postdoctoral
Fellowship grant nr. PF0-10011, awarded by the Chandra X-ray Center.
\end{acknowledgements}

\end{document}